\documentclass[10pt,conference,balance]{IEEEtran}
\usepackage{graphicx}  % Written by David Carlisle and Sebastian Rahtz
\usepackage{psfrag}    % Written by Craig Barratt, Michael C. Grant,
\usepackage{subfigure} % Written by Steven Douglas Cochran
\usepackage{amsmath}   % From the American Mathematical Society
\usepackage{amssymb}   % A popular package that provides many helpful commands
\usepackage{latexsym}
\usepackage{eucal}
\usepackage{balance}

\newtheorem{prop}{Proposition}
\newtheorem{conj}{Conjecture}

\renewcommand{\baselinestretch}{1.0}
\newcommand{\snr}{\rho}

\begin{document}

\title{\huge{Writing on Dirty Paper with Resizing and its Application to Quasi-Static Fading Broadcast Channels}}

\author{\authorblockN{Wenyi Zhang}
\authorblockA{Communication Science Institute\\
University of Southern California\\
Los Angeles, CA 90089\\
Email: wenyizha@usc.edu}
\and
\authorblockN{Shivaprasad Kotagiri and J. Nicholas Laneman}
\authorblockA{Department of Electrical Engineering\\
University of Notre Dame\\
Notre Dame, IN 46556\\
Email: \{skotagir, jlaneman\}@nd.edu}}

\maketitle

%%%%%%%%%%%%%%%%%%%%%%%%%%%%%%%%%%%%%%%%%%%%%%%%%%%%%%%%%%%%%%%%%%%%%%
%%%%%%%%%%%%%%%%%%%%%%%%%%%%%%%%%%%%%%%%%%%%%%%%%%%%%%%%%%%%%%%%%%%%%%
\begin{abstract}
This paper studies a variant of the classical problem of ``writing on dirty paper'' in which
the sum of the input and the interference, or dirt, is multiplied by a random variable that models resizing,
known to the decoder but not to the encoder. The achievable rate of Costa's
dirty paper coding (DPC) scheme is calculated and compared to the case of the decoder's also
knowing the dirt. In the ergodic case, the corresponding rate loss vanishes
asymptotically in the limits of both high and low signal-to-noise ratio (SNR),
and is small at all finite SNR for typical distributions like Rayleigh, Rician, and Nakagami.
In the quasi-static case, the DPC scheme is lossless at all SNR in terms of outage probability.
Quasi-static fading broadcast channels (BC) without transmit channel state information (CSI) are investigated
as an application of the robustness properties.
It is shown that the DPC scheme leads to an outage achievable rate region that strictly dominates that of time division.
\end{abstract}

%%%%%%%%%%%%%%%%%%%%%%%%%%%%%%%%%%%%%%%%%%%%%%%%%%%%%%%%%%%%%%%%%%%%%%
\section{Introduction}
\label{sec:intro}

In this paper we consider the following variant of Costa's ``writing on dirty paper'' (WDP) problem \cite{costa83:it},
\begin{eqnarray}
\label{eqn:wdp}
Y_i = \sqrt{A_i} \cdot (X_i + S_i) + Z_i,\quad i = 1, \ldots, n,
\end{eqnarray}
where the channel noise samples $\{Z_i\}_{i = 1}^n$ are independent, identically distributed (i.i.d.) circular-symmetric complex Gaussian with mean zero and variance $N$, {\it i.e.}, $Z_i \sim \mathcal{CN}(0, N)$ for $i = 1, \ldots, n$, and the interference signal samples $\{S_i\}_{i = 1}^n$ are i.i.d. $\mathcal{CN}(0, Q)$. The average power of the inputs $\{X_i\}_{i = 1}^n$ is constrained by
\begin{eqnarray*}
\frac{1}{n}\sum_{i = 1}^n |X_i|^2 \leq P.
\end{eqnarray*}
The non-negative random variables $\{A_i\}_{i = 1}^n$, with marginal probability density function (PDF) $p(a)$ for $a \in [0, \infty)$, model resizing coefficients or fading coefficients in wireless applications. Furthermore, $\{A_i\}_{i = 1}^n$ and $\{S_i\}_{i = 1}^n$ are independent.

We consider the following two scenarios:
\begin{itemize}
\item {\it Ergodic case:} $\{A_i\}_{i = 1}^n$ are $n$ i.i.d. random variables.
\item {\it Quasi-static case:} $\{A_i\}_{i = 1}^n$ remain constant over the entire coding block.
\end{itemize}

In both cases we assume that $\{A_i\}_{i = 1}^n$ are known to the decoder, but not to the encoder.
In the context of resizing and fading channels this is often a reasonable first-order approximation because the decoder
may estimate the coefficients with satisfactory accuracy, but the transmitter may not for lack of an adequate feedback link.
As in the original WDP problem, $\{S_i\}_{i = 1}^n$ are known to the encoder but not to the decoder.
In the sequel, for simplicity and without loss of generality, we drop the time index $i$, normalize $A$
such that it has unit expectation $\mathbf{E}[A] = 1$, and define the average signal-to-noise ratio
(SNR) as $\rho := {P}/{N}$.

When $A = 1$ with probability one, the problem reduces to the original WDP problem \cite{costa83:it}.
By introducing the auxiliary random variable
\begin{eqnarray}
U := X + \frac{\rho}{1 + \rho} S,
\end{eqnarray}
where the channel input $X \sim \mathcal{CN}(0, P)$ and is independent of the interference $S$, the capacity of the channel (\ref{eqn:wdp}) is given by \cite{costa83:it}
\begin{eqnarray}
C = I(U; Y) - I(U; S) = \log(1 + \rho).
\end{eqnarray}
The optimality of $C$ is apparent because it is lossless compared to the case in which the decoder knows $S$. In other words, the dirty paper coding (DPC) scheme can remove any effect of the additive interference.

The situation changes for general resizing  or fading scenarios in which the coefficient $A$
affects the realized or instantaneous SNR which is unknown to the encoder.
Therefore the encoder is incapable of dynamically adjusting its precoding to obtain optimality.

In the remainder of this paper, however, we demonstrate that the DPC scheme leads to fairly robust performance even for resizing or fading channels. We compare the achievable rate of the DPC scheme in our setup and the channel capacity when the decoder also knows the interference. In the ergodic case (Section \ref{sec:ergodic}), the DPC scheme turns out to be asymptotically lossless in the limits of both high and low SNR. Specifically, at high SNR the gap between the two rates vanishes, and at low SNR the ratio between the two rates approaches one. In the quasi-static case (Section \ref{sec:non-ergodic}), the DPC scheme is lossless at all SNR in terms of outage probability.

As one application of the revealed robustness properties, we investigate quasi-static fading broadcast channels (BC) without transmit channel state information (CSI) (Section \ref{sec:bc}). It is shown that the DPC scheme leads to an outage achievable rate region equivalent to the capacity region of a corresponding BC without fading. This rate region is, to the best of the authors' knowledge, the first that strictly dominates that of time division.

%%%%%%%%%%%%%%%%%%%%%%%%%%%%%%%%%%%%%%%%%%%%%%%%%%%%%%%%%%%%%%%%%%%%%%
\section{Ergodic Case}
\label{sec:ergodic}

\subsection{Some Preliminaries}
\label{sec:background}

From a high-level perspective, the ergodic case falls into the category of channel coding with two-sided state information \cite{cover02:it}. In \cite[Theorem 1]{cover02:it} the authors consider a discrete memoryless channel $p(y|x, s_1, s_2)$ with state information $(S_{1, i}, S_{2, i})$ i.i.d. $\sim p(s_1, s_2)$, $S_1$ non-causally known to the encoder, and $S_2$ known to the decoder.\footnote{Since we always consider block decoding, it makes no difference whether we assume causal or non-causal state information at the decoder.} They show that the channel has capacity
\begin{eqnarray}
\label{eqn:general-formula}
C = \max_{p(u, x|s_1)} \left[I(U; S_2, Y) - I(U; S_1)\right].
\end{eqnarray}

We note that the proof of (\ref{eqn:general-formula}) in \cite{cover02:it} requires that the alphabets $|\mathcal{X}|$, $|\mathcal{S}_1|$, $|\mathcal{S}_2|$, $|\mathcal{Y}|$ be finite. When this condition is violated, for example in the WDP problem, (\ref{eqn:general-formula}) provides an achievable rate, but it remains unclear if it also provides the converse. In this paper, we focus on utilizing (\ref{eqn:general-formula}) to establish achievable rate for the particular class of $U$ generated by the DPC scheme.

As a side note, (\ref{eqn:general-formula}) may also be interpreted by treating $S_2$ as an additional channel output,
and adopting an argument similar to that in \cite{caire99:it} for channels with causal state information.

We adopt the DPC scheme of \cite{costa83:it}. The channel input $X$ is i.i.d. $\mathcal{CN}(0, P)$, and
the auxiliary random variable $U = X + \alpha S$, where $\alpha$ is a constant to be designed.
Noting that the coefficient $A$ is independent of $U$, we have an achievable rate
\begin{eqnarray}
\label{eqn:general}
R = I(U; A, Y) - I(U; S) = I(U; Y|A) - I(U; S).
\end{eqnarray}
Following algebraic manipulations similar to those in \cite{costa83:it}, we obtain
\begin{eqnarray}
\label{eqn:general-rate}
R = \mathbf{E}\left[
\log \frac{P [A(P + Q) + N]}{(1 - \alpha)^2 APQ + (P + \alpha^2 Q)N}
\right],
\end{eqnarray}
where the expectation is with respect to the distribution $p(a)$ of $A$.

For every given configuration of $(P, Q, N)$ and $p(a)$, maximizing (\ref{eqn:general-rate}) gives the optimal choice of $\alpha$
and the corresponding maximized $R$. In the following, however, we focus on the particular choice of $\alpha = \rho/(\rho + 1)$.
That is, even though the channel is time-varying, we precode for the {\it average} SNR rather than the instantaneous
SNR for each channel use.

For simplicity we define the interference-to-power ratio (IPR) $\beta := {Q}/{P}$. As we let $\alpha = \rho/(\rho + 1)$, (\ref{eqn:general-rate}) becomes
\begin{eqnarray}
\label{eqn:rate-para}
R = \mathbf{E}\left[
\log \frac{(\rho + 1)^2 \left[(1 + \beta)A\rho + 1\right]}{(\beta + 1)\rho^2 + (\beta A + 2)\rho + 1}
\right].
\end{eqnarray}

When the decoder also knows the interference $S$, it can simply subtract off $\sqrt{A}S$, and the capacity is
\begin{eqnarray}
\label{eqn:cap-noint}
\bar{C} = \mathbf{E}[\log (1 + A\rho)],
\end{eqnarray}
which provides a performance upper bound for comparison in the following.

\subsection{Asymptotics}

{\it (1) High SNR:} As $\rho \rightarrow \infty$, by expanding (\ref{eqn:rate-para}) with respect to $1/\rho$, we obtain
\begin{eqnarray}
\label{eqn:high-snr}
R = \log \rho + \mathbf{E}[\log A] + \frac{\beta + \mathbf{E}[1/A]}{\beta + 1} \cdot (1/\rho) + o(1/\rho).
\end{eqnarray}
Comparing (\ref{eqn:high-snr}) with the high-SNR expansion of (\ref{eqn:cap-noint})
\begin{eqnarray}
\label{eqn:high-snr-reg}
\bar{C} = \log \rho + \mathbf{E}[\log A] + \mathbf{E}[\frac{1}{A}] \cdot (1/\rho) + o(1/\rho),
\end{eqnarray}
we observe that their difference vanishes as $\rho \rightarrow \infty$.\footnote{The expansions of (\ref{eqn:high-snr}) and (\ref{eqn:high-snr-reg}) implicitly assume that $\mathbf{E}[1/A] < \infty$. When this is not the case, these expansions involve fractional-order terms, but the conclusion that their difference vanishes as $\rho \rightarrow \infty$ can still be shown to hold.}

{\it (2) Low SNR:} As $\rho \rightarrow 0$, by expanding (\ref{eqn:rate-para}) with respect to $\rho$, we obtain
\begin{eqnarray}
\label{eqn:low-snr}
R = \rho - \left[
(2\beta + 1)\mathbf{E}[A^2] - 2\beta
\right] \cdot\rho^2 + o(\rho^2).
\end{eqnarray}
Comparing (\ref{eqn:low-snr}) with the low-SNR expansion of (\ref{eqn:cap-noint})
\begin{eqnarray}
\bar{C} = \rho - \mathbf{E}[A^2]\rho^2 + o(\rho^2),
\end{eqnarray}
we observe that their ratio approaches one as $\rho \rightarrow 0$.

\subsection{Finite SNR Behavior}

For finite SNR, we denote the gap between $\bar{C}$ and $R$ by $\Delta$, which is
\begin{eqnarray*}
\Delta &:=& \bar{C} - R\nonumber\\
&=& \mathbf{E}\left[
\log\frac{(A\rho + 1) [
(\beta + 1)\rho^2 + (\beta A + 2)\rho + 1
]}{(\rho + 1)^2 [(\beta + 1)A\rho + 1]}
\right].
\end{eqnarray*}
An upper bound to $\Delta$ is easily obtained by letting $\beta \rightarrow \infty$:
\begin{eqnarray}
\label{eqn:bar-delta}
\Delta \leq \bar{\Delta} := \mathbf{E}\left[
\log (\rho + A) + \log (\rho + 1/A)
\right] - 2 \log (\rho + 1).
\end{eqnarray}

From (\ref{eqn:bar-delta}), we can evaluate $\bar{\Delta}$ either analytically or numerically using Monte Carlo, thereby obtaining an upper bound to the possible rate loss of the DPC scheme. For Rayleigh fading channels, it is found that $\bar{\Delta}$ is maximized around $\rho = 0$ dB, and the maximum value is about $0.384$. That is, the performance loss due to the DPC scheme is less than $0.4$ nats per channel use, at all SNR. Similarly, we obtain numerical results for Rician and Nakagami fading channels. For Rician fading channels, the maximum $\bar{\Delta}$ across SNR never exceeds that when the line-of-sight (LOS) component is absent, {\it i.e.}, when the channel is Rayleigh; and is further reduced when the LOS component increases, {\it i.e.}, when the channel becomes more Gaussian. For Nakagami channels,
the maximum $\bar{\Delta}$ monotonically decreases as the fading figure $m \geq 1/2$  increases \cite{proakis95:book}, and never exceeds one nat per channel use.

%%%%%%%%%%%%%%%%%%%%%%%%%%%%%%%%%%%%%%%%%%%%%%%%%%%%%%%%%%%%%%%%%%%%%%
\section{Quasi-Static Case}
\label{sec:non-ergodic}

In the quasi-static fading case, the coefficient $A$ is one realization drawn according to $p(a)$. Using the DPC scheme with $\alpha$ independent of $A$, from \cite{costa83:it} we have that the achievable rate is the random variable
\begin{eqnarray}
\label{eqn:cond-rate}
J(\alpha, A) :=
\log \frac{P[A(P + Q) + N]}{(1 - \alpha)^2 APQ + (P + \alpha^2 Q)N},
\end{eqnarray}
induced by $A$. For every $R \geq 0$, we can calculate the outage probability $\mathrm{Pr}\left[A: J(\alpha, A) \leq R\right]$, {\it i.e.}, the probability that the realization of $A$ makes the achievable rate $J(\alpha, A)$ insufficient to support the target rate $R$. Furthermore, let us adjust $\alpha$ to minimize the outage probability for every given $R$. To this end we notice that
\begin{eqnarray}
&&\mathrm{Pr}\left[A: J(\alpha, A) \leq R\right] \nonumber\\
&=& \mathrm{Pr}\left[
A: A \leq \frac{(e^{R} - 1)PN + \alpha^2 e^{R} QN}{P\left(
P + Q - (1 - \alpha)^2 e^{R} Q
\right)}
\right].
\end{eqnarray}
Hence by straightforward manipulations, we find that the minimizer of $\mathrm{Pr}\left[A: J(\alpha, A) \leq R\right]$ is
%\begin{eqnarray}
%\label{eqn:alpha-opt}
$\alpha^\ast = 1 - e^{-R}$,
%\end{eqnarray}
and that the corresponding minimum outage probability is
\begin{eqnarray}
\label{eqn:min-outage}
\min \mathrm{Pr}\left[A: J(\alpha, A) \leq R\right] = \mathrm{Pr}\left[
A: A \leq \frac{e^{R} - 1}{\rho}
\right],
\end{eqnarray}
where $\rho := P/N$ as in the previous sections.

{\it Discussion:}

(1) In view of (\ref{eqn:min-outage}), we observe that this minimum outage probability coincides with the minimum outage probability when the decoder also knows $S$. That is, in the quasi-static case, the DPC scheme is optimal at all SNR, regardless of the specific distribution for $A$. Furthermore, it is important to note that the optimal choice of $\alpha$ depends upon the target rate $R$. In fact, we may introduce a virtual SNR $\rho^\ast$ satisfying $R = \log(1 + \rho^\ast)$, and rewrite $\alpha^\ast = 1 - e^{-R} = {\rho^\ast}/{(1 + \rho^\ast)}$. That is, for a given target rate, the optimal strategy is to simply treat the channel as if it is realized to just be able to support this rate.

Such behavior can be explained by the following coincidence argument. The conditional achievable rate $J(\alpha, A)$ as given by (\ref{eqn:cond-rate}) is a function of two variables, $\alpha$ and $A$. It can be verified that for every $\alpha$, $J(\alpha, A)$ is monotonically increasing with $A$. On the other hand, the DPC scheme shows that, for every $A$, the optimal choice of $\alpha$ maximizing $J(\alpha, A)$ is given by $\alpha^{\mathrm{DPC}}(A) = A \rho/(A \rho + 1)$. Therefore, for a given target rate $R$, if we solve $J(\alpha^{\mathrm{DPC}}(A), A) = R$ for its solution $A^\ast$ and choose $\alpha^\ast = \alpha^\mathrm{DPC}(A^\ast)$ in the DPC scheme, we can guarantee that for every fading realization $A < A^\ast$, the target rate $R$ is achievable.

\begin{figure}[h]
\begin{center}
\psfrag{x-label}{{$J(\alpha, A)$}}
\psfrag{y-label}{{CDF}}
\resizebox{0.9\columnwidth}{!}{
\includegraphics{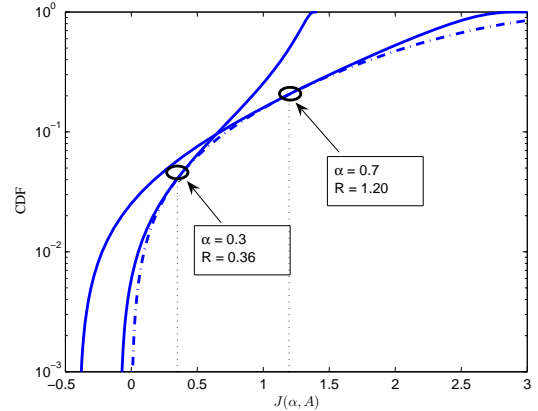}}
\end{center}
\caption{The CDF of $J(\alpha, A)$ for Rayleigh fading channels.} \label{fig:robustness}
\end{figure}

(2) To illustrate how the DPC scheme achieves the same outage probability as if there were no interference, we can plot the cumulative distribution function (CDF) of $J(\alpha, A)$. Assuming that the fading is Rayleigh such that $A$ follows the exponential distribution with unit mean, Figure \ref{fig:robustness} displays the CDF for $P/N = 10$ and $Q = P$. Plotted are two solid curves corresponding to $\alpha = 0.3$ with $R \approx 0.36$ nats and $\alpha = 0.7$ with $R \approx 1.20$ nats, respectively.
It is clear that the CDF of the DPC scheme depends upon the particular choice of $\alpha$.
The reference CDF, with $Q = 0$, is plotted as the dashed-dot curve.
It is clearly illustrated that, for a given target rate $R$, if we choose the corresponding $\alpha^\ast = 1 - e^{-R}$, the CDF of the DPC scheme tangentially intersects the reference CDF exactly at $J(\alpha, A) = R$, hence resulting in the identical outage probability.

% \begin{figure}[t]
% \psfrag{x-label}{\small{$J(\alpha, A)$}}
% \psfrag{y-label}{\small{CDF}}
% \epsfxsize=3.3in
% \epsfclipon
% \centerline{\epsffile{./robustness.eps}}
% \caption{The CDF of $J(\alpha, A)$ for Rayleigh fading channels.}
% \label{fig:robustness}
% \end{figure}

(3) Finally we note that the conditional achievable rate $J(\alpha, A)$ is negative for a certain range of $A$, as calculated from (\ref{eqn:cond-rate}). This implies that for some values of $A$, the DPC scheme with the choice of $\alpha$ cannot lead to nonnegative achievable rate. From a practical perspective, whenever $J(\alpha, A) \leq 0$, it loses no generality to replace it with $J(\alpha, A) = 0$.

%%%%%%%%%%%%%%%%%%%%%%%%%%%%%%%%%%%%%%%%%%%%%%%%%%%%%%%%%%%%%%%%%%%%%%
\section{Application to Quasi-Static Fading BC}
\label{sec:bc}

If we treat the interference $S$ in the channel model (\ref{eqn:wdp}) as a source image, the robustness results obtained in the preceding sections can be immediately applied in digital watermarking with resizing or  transmission over fading channels.
Alternatively, we can treat both the signal $X$ and the interference $S$ as coded messages, then such a two-layer
coding scheme can be useful in the fading BC. In this section, we illustrate that the robustness property
of the DPC scheme can be useful for quasi-static fading BC without transmit CSI.

\subsection{Background Overview}

The Gaussian BC has been investigated since the seminal paper of Cover in 1972 \cite{cover72:it}.
It is shown that for scalar Gaussian BC, superposition coding achieves a rate region that dominates that of time division,
and actually yields the capacity region \cite{bergmans74:it-1}.
If the channel inputs are vector-valued, the vector Gaussian BC is generally non-degraded,
 and superposition coding ceases to yield optimal performance. On the other hand, the DPC scheme has been
 shown to maximize the throughput of the Gaussian vector BC \cite{caire03:it}. This observation has stimulated a
 series of subsequent work on multi-antenna BC \cite{yu04:it, viswanath02:dimacs, vishwanath03:it}. Recently the achievable rate region obtained by the DPC scheme in \cite{caire03:it} has been shown to be the capacity region \cite{weingarten06:it}.

A central prerequisite of the aforementioned results is that the encoder has full access to CSI. It is shown in \cite{lapidoth05:allerton} that, if the transmit CSI is noisy, then the high SNR growth rate of the channel throughput is significantly reduced compared to that for perfect transmit CSI. For fading BC with ergodic fading, when the encoder possesses no CSI, neither the capacity region nor the maximum channel throughput is known. Some achievable rate regions are obtained in \cite{tuninetti03:isit, jafar05:it}, and it is conjectured therein that Gaussian inputs achieve the capacity region.

The requirement of transmit CSI appears even more crucial for quasi-static channels. %For slowly time-varying applications, it customary to adopt a quasi-static fading model, which exhibits a non-ergodic characteristic with the fading process remaining constant over a whole coding block, and independently changing to a new realization in the next coding block.
Consider a scalar BC. When transmit CSI is available, the encoder then knows exactly which decoder is degraded, possibly changing from one coding block to another. Superposition coding therefore can be utilized, such that in each coding block the ``less noisy'' decoder performs successive decoding, and the resulting achievable rate region is the capacity region conditioned upon the fading realization. Consequently, the outage capacity region can be identified \cite{li01:it}.

However, if transmit CSI is absent, little progress has been reported. The main difficulty appears to be that the lack of transmit CSI prevents the encoder from making efficient utilization of superposition coding, because there is generally no intelligent way to identify which decoder is less noisy, even for the scalar case. Generally, the quasi-static fading BC without transmit CSI is non-degraded, and belongs to the class of mixed channels as defined in \cite{han03:book}. For such channels, no computable single-letter characterization of their $\epsilon$-capacity regions, {\it i.e.}, outage capacity regions, has been identified \cite{iwata05:ieice}.

\subsection{Two-User Case}

Consider the two-user quasi-static fading BC model in which
\begin{eqnarray}
Y_i = \sqrt{A_i} X + Z_i,\quad i = 1, 2.
\end{eqnarray}
The additive noise $Z_i \sim \mathcal{CN}(0, 1)$, and the input $X$ has an average power constraint $\snr$. The fading coefficient $A_i$ is known at the corresponding decoder $i$, but not at the encoder. We assume that $A_i$ has PDF $p_i(a)$ for $a \in [0, \infty)$
and that $(A_1, A_2)$ are independent and remain constant over the entire coding block. In the sequel, the inverse cumulative distribution function (ICDF) of $p_i(\cdot)$, denoted by $G_i(\cdot)$, will be of use. For every $t \in [0, 1]$, $G_i(t)$ is defined as the infimum such that $\mathrm{Pr}[A_i \leq G_i(t)] = t$.

%The achievable rate region of the two-user quasi-static fading BC is a random two-dimensional set, as induced by the fading coefficients $\mathbf{A} := (A_1, A_2)$. Therefore,
For any given target outage probability vector $\underline{\epsilon} := ({\epsilon}_1, {\epsilon}_2) \in [0, 1]^2$, we can define the $\underline{\epsilon}$-outage capacity region $\mathcal{C}^\mathrm{out}(\snr, \underline{\epsilon})$ as the union of all rate vectors $\mathbf{R} = (R_1, R_2)$ that can be achieved with the outage probability for decoder $i$ no larger than ${\epsilon}_i$, $i = 1, 2$ \cite{li01:it}. To simplify the notation and discussion,  we assume $G_1(\epsilon_1) \geq G_2(\epsilon_2)$.

A conceptually simple scheme is time-division (TD) coding, with time-sharing factor $(\mu, 1 - \mu)$ and power allocation factor $(\eta_1, \eta_2)$. This yields the boundary of the $\underline{\epsilon}$-outage achievable rate region
\begin{equation}
\mathcal{R}^\mathrm{tdpa}(\snr, \underline{\epsilon}) := \bigcup_{\stackrel{\mu \in [0, 1]}{\eta_i \geq 0, i = 1,2}} \left(R^\mathrm{tdpa}_1(\snr, \mu, \eta_1, \underline{\epsilon}), R^\mathrm{tdpa}_2(\snr, \mu, \eta_2, \underline{\epsilon})\right),
\end{equation}
where
\begin{eqnarray*}
R^\mathrm{tdpa}_1(\snr, \mu, \eta_1, \underline{\epsilon}) &:=& \mu\log[1 + \eta_1 G_1({\epsilon}_1) \cdot\snr]\\
R^\mathrm{tdpa}_2(\snr, \mu, \eta_2, \underline{\epsilon}) &:=& (1 - \mu)\log[1 + \eta_2 G_2({\epsilon}_2) \cdot\snr]\\
\mu \eta_1 + (1 - \mu) \eta_2 &=& 1.
\end{eqnarray*}

It turns out that we can do better than $\mathcal{R}^\mathrm{tdpa}(\snr, \underline{\epsilon})$, if we appropriately utilize the robustness property of DPC as revealed in Section \ref{sec:non-ergodic}. We split the input into $X = X_1 + X_2$, with $X_1 \sim \mathcal{CN}(0, \gamma \snr)$, $X_2 \sim \mathcal{CN}(0, (1 - \gamma)\snr)$, and independent.
The message for decoder 2 is encoded as $X_2$ simply chosen from a Gaussian codebook and the message for decoder 1
is encoded as $X_1$ treating $X_2$ as interference using the DPC scheme.

From Section \ref{sec:non-ergodic}, for target rate vector $\mathbf{R} = (R_1, R_2)$, decoder 1 experiences outage if
\begin{eqnarray*}
A_1 \leq \frac{\exp(R_1) - 1}{\gamma \snr} \;\mathrm{or} \; \log(1 + \gamma A_1 \snr) \leq R_1.
\end{eqnarray*}
On the other hand, decoder 2 treats $X_1$ as additive noise, therefore it experiences outage if
\begin{equation*}
\log\left(
\frac{A_2 \snr + 1}{A_2 \gamma \snr + 1}
\right) \leq R_2.
\end{equation*}
So the boundary of the $\underline{\epsilon}$-outage achievable rate region for the DPC scheme is
\begin{equation}
\mathcal{R}^\mathrm{dpc}(\snr, \underline{\epsilon}) := \bigcup_{\gamma \in [0, 1]} \left(R^\mathrm{dpc}_1(\snr, \gamma, \underline{\epsilon}), R^\mathrm{dpc}_2(\snr, \gamma, \underline{\epsilon})\right),
\end{equation}
where
\begin{eqnarray*}
R^\mathrm{dpc}_1(\snr, \gamma, \underline{\epsilon}) &:=& \log[1 + \gamma G_1({\epsilon}_1) \snr]\\
R^\mathrm{dpc}_2(\snr, \gamma, \underline{\epsilon}) &:=& \log\left(1 + \frac{(1-\gamma)G_2({\epsilon}_2) \snr}{1 + \gamma G_2({\epsilon}_2) \snr}\right).
\end{eqnarray*}

We observe that
$\mathcal{R}^\mathrm{tdpa}(\snr, \underline{\epsilon})$ and $\mathcal{R}^\mathrm{dpc}(\snr, \underline{\epsilon})$
can also be interpreted as achievable rate regions of a BC without fading in which
\begin{eqnarray}
\tilde{Y}_i = \sqrt{G_i({\epsilon}_i)} \tilde{X} + \tilde{Z}_i,\quad i = 1, 2.
\end{eqnarray}
Specifically, $\mathcal{R}^\mathrm{tdpa}(\snr, \underline{\epsilon})$ is achieved by time-division
schemes with power allocation, and since
$G_1({\epsilon}_1) \geq G_2({\epsilon}_2)$, $\mathcal{R}^\mathrm{dpc}(\snr, \underline{\epsilon})$
is achieved by superposition coding and therefore is the capacity region.
Consequently, we have that whenever $G_1({\epsilon}_1) \geq G_2({\epsilon}_2)$,
the $\underline{\epsilon}$-outage achievable rate region of the DPC scheme,
$\mathcal{R}^\mathrm{dpc}(\snr, \underline{\epsilon})$, contains that of time division,
$\mathcal{R}^\mathrm{tdpa}(\snr, \underline{\epsilon})$  \cite{bergmans74:it-2}.

%As a side note, the DPC scheme also yields an achievable rate region even if $G_1({\epsilon}_1) < G_2({\epsilon}_2)$. This case is less %interesting, because it can be shown that $\mathcal{R}^\mathrm{dpc}(\snr, \underline{\epsilon})$ for $G_1({\epsilon}_1) < G_2({\epsilon}_2)$ %is always dominated by $\mathcal{R}^\mathrm{tdpa}(\snr, \underline{\epsilon})$.

\subsection{General $K$-User Case}

We can extend the two-user BC result to the general $K$-user case. The channel model is
\begin{eqnarray}
\label{eqn:k-user}
Y_k = \sqrt{A_k} X + Z_k,\quad\mathrm{for}\;k = 1, \ldots, K,
\end{eqnarray}
where $Z_k \sim \mathcal{CN}(0, 1)$, and $X$ has an average power constraint $\snr$.
Similarly, each $A_k$, following PDF $p_k(a)$ for $a \in [0, \infty)$ and remaining constant over the entire coding block,
is known at decoder $k$ but not at the encoder. The $A_k$'s are mutually independent.  We have the following result.
\begin{prop} \label{prop:k-user}

For the $K$-user quasi-static scalar fading BC without transmit CSI, (\ref{eqn:k-user}), the DPC scheme achieves the boundary of the $\underline{\epsilon}$-outage achievable rate region
\begin{eqnarray}
\label{eqn:R-dpc}
\mathcal{R}^\mathrm{dpc}(\snr, \underline{\epsilon}) := \bigcup_{\underline{\gamma}} \left(R^\mathrm{dpc}_1(\snr, \underline{\gamma}, \epsilon_1), \ldots, R^\mathrm{dpc}_K(\snr, \underline{\gamma}, \epsilon_K)\right),
\end{eqnarray}
where
\begin{eqnarray*}
\label{eqn:R-dpc-k}
R^\mathrm{dpc}_k(\snr, \underline{\gamma}, \epsilon_k) &:=& \log\left(1 +
\frac{\gamma_k G_k({\epsilon}_k) \cdot \snr}
{(\sum_{i = 1}^{k - 1}\gamma_i) G_k({\epsilon}_k)\cdot \snr + 1}
\right),\\%\quad\mathrm{for}\;k = 1, \ldots, K\\
\mathrm{and}\;\sum_{k = 1}^K \gamma_k &=& 1,\;\mathrm{for}\; \gamma_k \geq 0, k = 1, \ldots, K.
\end{eqnarray*}
% The boundary $\mathcal{R}^\mathrm{dpc}(\snr, \underline{\epsilon})$ is maximized by ordering the $K$ decoders such that $G_1(\epsilon_1) \geq G_2(\epsilon_2) \geq \ldots \geq G_K(\epsilon_K)$.
\end{prop}

In view of $\mathcal{R}^\mathrm{dpc}(\snr, \underline{\epsilon})$, we again observe that if $G_1(\epsilon_1) \geq G_2(\epsilon_2) \geq \ldots \geq G_K(\epsilon_K)$, then $\mathcal{R}^\mathrm{dpc}(\snr, \underline{\epsilon})$ is exactly the capacity region of the BC without fading
\begin{eqnarray}
\tilde{Y}_k = \sqrt{G_k({\epsilon}_k)} \tilde{X} + \tilde{Z}_k,\quad \mathrm{for}\;k = 1, \ldots, K.
\end{eqnarray}
In light of this observation, we conjecture that the DPC scheme with appropriate ordering of the encoders
 is optimal in terms of the outage achievable rate region.
\begin{conj}
For the $K$-user quasi-static scalar fading BC without transmit CSI, (\ref{eqn:k-user}), if the $K$ decoders are ordered such that $G_1(\epsilon_1) \geq G_2(\epsilon_2) \geq \ldots \geq G_K(\epsilon_K)$, then $\mathcal{R}^\mathrm{dpc}(\snr, \underline{\epsilon})$ is the boundary of the $\underline{\epsilon}$-outage capacity region $\mathcal{C}^\mathrm{out}(\snr, \underline{\epsilon})$.
\end{conj}

{\it Proof of Proposition \ref{prop:k-user}:} The proof essentially follows the same idea of the two-user case.
Fix the power allocation $\underline{\gamma}$. For decoder $k$, rewrite the channel as
\begin{eqnarray}
\label{eqn:faded-noise}
Y_k = \sqrt{A_k} X_k + \sqrt{A_k} \sum_{i > k} X_i + (\sqrt{A_k} \sum_{j < k} X_j + Z_k),
\end{eqnarray}
where $X_k$ denotes the encoded signal for decoder $k$. Let $X_k$ be mutually independent and
$X_k \sim \mathcal{CN}(0, \gamma_i \snr)$ for $k = 1, \ldots, K$. The message for decoder $k$
is encoded as $X_k$, by the DPC scheme treating $\sum_{i > k} X_i$ as transmit interference, and
treating $(\sqrt{A_k} \sum_{j < k} X_j + Z_k)$ as noise.

Following the same coincidence argument as in Section \ref{sec:non-ergodic}, we can extend the
robustness property of the DPC scheme to the channel (\ref{eqn:faded-noise}).
That is, for target rate $R_k$, if we choose $\alpha^\ast_k = 1 - e^{-R_k}$ in the DPC scheme,
the outage probability is $\mathrm{Pr}\left[
A_k: A_k \leq \frac{e^{R_k} - 1}{\gamma_k \snr - (e^{R_k} - 1) (\sum_{j < k}\gamma_j) \snr}
\right]$, as if the transmit interference $\sum_{i > k} X_i$ is known at the decoder $k$.
Equivalently, for a given target outage probability $\epsilon_k$ for decoder $k$,
the maximum achievable rate $R_k$ should satisfy
\begin{eqnarray*}
\frac{e^{R_k} - 1}{\gamma_k \snr - (e^{R_k} - 1) (\sum_{j < k}\gamma_j) \snr} = G_k(\epsilon_k),
\end{eqnarray*}
which immediately gives rise to one point of the boundary
$\mathcal{R}^\mathrm{dpc}(\underline{\epsilon})$ for the fixed power allocation $\underline{\gamma}$.
The entire boundary $\mathcal{R}^\mathrm{dpc}(\underline{\epsilon})$ then follows,
as we exhaust all the possible choices of $\underline{\gamma}$.

%%%%%%%%%%%%%%%%%%%%%%%%%%%%%%%%%%%%%%%%%%%%%%%%%%%%%%%%%%%%%%%%%%%%%%%%%%%
%%%%%%%%%%%%%%%%%%%%%%%%%%%%%%%%%%%%%%%%%%%%%%%%%%%%%%%%%%%%%%%%%%%%%%%%%%%
\section*{Acknowledgment}
The work of W. Zhang has been supported in part by NSF NRT ANI-0335302,
NSF ITR CCF-0313392, and NSF OCE-0520324.
The work of S. Kotagiri and J. N. Laneman has been supported in
part by NSF CCF05-46618, and a fellowship from
the University of Notre Dame Center for Applied Mathematics. W. Zhang wishes to thank
Giuseppe Caire for encouragement and useful comments in writing this paper.

%%%%%%%%%%%%%%%%%%%%%%%%%%%%%%%%%%%%%%%%%%%%%%%%%%%%%%%%%%%%%%%%%%%%%%
%%%%%%%%%%%%%%%%%%%%%%%%%%%%%%%%%%%%%%%%%%%%%%%%%%%%%%%%%%%%%%%%%%%%%%
\balance
\bibliographystyle{ieee}
\bibliography{../../thesis/thesis}

\begin{thebibliography}{1}

\bibitem{costa83:it}
M.~H.~M. Costa,
\newblock ``{Writing on Dirty Paper},''
\newblock {\em \textit{{IEEE} Trans. Inform. Theory}}, vol. 29, no. 3, pp.
  439--441, May 1983.
%
%\bibitem{caire98:it}
%Guiseppe Caire, Giorgio Taricco, and Ezio Biglieri,
%\newblock ``{Bit-Interleaved Coded Modulation},''
%\newblock {\em \textit{{IEEE} Trans. Inform. Theory}}, vol. 44, no. 3, pp.
%  927--946, May 1998.

\bibitem{cover02:it}
T.~M. Cover and M. Chiang,
\newblock ``{Duality Between Channel Capacity and Rate Distortion With
  Two-Sided State Information},''
\newblock {\em \textit{{IEEE} Trans. Inform. Theory}}, vol. 48, no. 6, pp.
  1629--1638, Jun. 2002.
  
\bibitem{caire99:it}
G. Caire and S. Shamai (Shitz),
\newblock ``{On the Capacity of Some Channels with Channel State Information},''
\newblock {\em \textit{{IEEE} Trans. Inform. Theory}}, vol. 45, no. 6, pp. 2007--2019, Sep. 1999.

\bibitem{proakis95:book}
J.~G. Proakis,
\newblock {\em {Digital Communications}},
\newblock McGraw-Hill, Inc., New York, third edition, 1995.

%\bibitem{cohen02:it}
%A.~S. Cohen and A. Lapidoth,
%\newblock ``{The Gaussian Watermarking Game},''
%\newblock {\em \textit{{IEEE} Trans. Inform. Theory}}, vol. 48, no. 6, pp. 1639--1667, Jun. 2002.

\bibitem{cover72:it}
T. M. Cover,
\newblock ``Broadcast Channels,''
\newblock {\em \textit{IEEE Trans. Inform. Theory}}, vol. 18, no. 1, pp. 2--14, Jan. 1972.

\bibitem{bergmans74:it-1}
P. P. Bergmans,
\newblock ``A Simple Converse for Broadcast Channels with Additive White Gaussian Noise,''
\newblock {\em \textit{IEEE Trans. Inform. Theory}}, vol. 20, no. 2, pp. 279--280, Mar. 1974.

\bibitem{caire03:it}
G. Caire and S. Shamai (Shitz),
\newblock ``{On the Achievable Throughput of a Multi-antenna Gaussian Broadcast Channel},''
\newblock {\em \textit{{IEEE} Trans. Inform. Theory}}, vol. 49, no. 7, pp. 1691--1706, Jul. 2003.

\bibitem{yu04:it}
W. Yu and J. M. Cioffi,
\newblock ``Sum Capacity of Gaussian Vector Broadcast Channels,''
\newblock {\em \textit{{IEEE} Trans. Inform. Theory}}, vol. 50, no. 9, pp. 1875--1892, Sep. 2004.

\bibitem{viswanath02:dimacs}
P. Viswanath and D. N. C. Tse,
\newblock ``On the Capacity of the Multiple Antenna Broadcast Channel,''
\newblock {\em \textit{DIMACS Workshop on Signal Processing for Wireless Communications}}, Oct. 2002.

\bibitem{vishwanath03:it}
S. Vishwanath, N. Jindal, and A. Goldsmith,
\newblock ``Duality, Achievable Rates, and Sum-Rate Capacity of Gaussian MIMO Broadcast Channels,''
\newblock {\em \textit{{IEEE} Trans. Inform. Theory}}, vol. 49, no. 10, pp. 2658--2668, Oct. 2003.

\bibitem{weingarten06:it}
H. Weingarten, Y. Steinberg, and S. Shamai (Shitz),
\newblock ``The Capacity Region of the Gaussian Multiple-Input Multiple-Output Broadcast Channel,''
\newblock {\em \textit{{IEEE} Trans. Inform. Theory}}, vol. 52, no. 9, pp. 3936--3964, Sep. 2006.

\bibitem{lapidoth05:allerton}
A. Lapidoth, S. Shamai (Shitz), and M. A. Wigger,
\newblock ``On the Capacity of Fading MIMO Broadcast Channels with Imperfect Transmitter Side-Information,''
\newblock {\em in 43rd Allerton Conference on Communications, Control, and Computing}, Monticello, IL, USA, 2005.

\bibitem{tuninetti03:isit}
D. Tuninetti and S. Shamai (Shitz),
\newblock ``The Capacity Region of Two User Fading Broadcast Channels with Perfect Channel State Information at the Receivers,''
\newblock {\em in Proc. IEEE Int. Symp. Inform. Theory (ISIT)}, Yokohama, Japan, 2003.

\bibitem{jafar05:it}
S. A. Jafar and A. Goldsmith,
\newblock ``Isotropic Fading Vector Broadcast Channels: The Scalar Upperbound and Loss in Degrees of Freedom,''
\newblock {\em \textit{{IEEE} Trans. Inform. Theory}}, vol. 51, no. 3, pp. 848--857, Mar. 2005.

\bibitem{li01:it}
L. Li and A. J. Goldsmith,
\newblock ``{Capacity and Optimal Resource Allocation for Fading Broadcast Channels -- Part II: Outage Capacity},''
\newblock {\em \textit{{IEEE} Trans. Inform. Theory}}, vol. 47, no. 3, pp. 1103--1127, Mar. 2001.

\bibitem{han03:book}
T. S. Han,
\newblock {\em Information-Spectrum Methods in Information Theory},
\newblock Springer-Verlag, Berlin, 2003.

\bibitem{iwata05:ieice}
K-i Iwata and Y. Oohama,
\newblock ``Information-Spectrum Characterization of Broadcast Channel with General Source,''
\newblock {\em \textit{IEICE Trans. Fundamentals}}, vol. E88-A, no. 10, pp. 2808--2818, Oct. 2005.

\bibitem{bergmans74:it-2}
P. P. Bergmans and T. M. Cover,
\newblock ``Cooperative Broadcasting,''
\newblock  {\em \textit{{IEEE} Trans. Inform. Theory}}, vol. 20, no. 3, pp. 317--324, May 1974.

%\bibitem{gelfand80:pcit}
%S. I. Gel'fand and M. S. Pinsker,
%\newblock ``Coding for Channel with Random Parameters,''
%\newblock {\em \textit{Problems of Control and Information Theory}}, vol. 9, no. 1, pp. 19--31, 1980.

\end{thebibliography}

\end{document}